\documentclass[twocolumn,floats,floatfix,superscriptaddress,aps,pra]{revtex4-2}

\usepackage{lineno}

\usepackage{amsfonts}
\usepackage{amssymb}
\usepackage{amsmath}
\usepackage{bm,siunitx}
\usepackage{color}
\usepackage{epsfig}
\usepackage{ifthen}
\usepackage{graphicx}
\usepackage{multirow}
\usepackage{bigstrut}
\usepackage{rotating} 
\usepackage{soul}
\usepackage{xcolor}
\usepackage{comment}
\usepackage{subcaption}
\usepackage{caption}
\sethlcolor{yellow}

\def\be{ \begin{equation} }
\def\ee{ \end{equation} }
\def\bea{ \begin{eqnarray} }
\def\eea{ \end{eqnarray} }
\def\bse{ \begin{subequations} }
\def\ese{ \end{subequations} }

\def\P01{P_{\,\text{0}\rightarrow \text{1}} }

\def\bt{\begin{tabular}}
\def\et{\end{tabular}}

\def\O{\Omega}

\begin{document}

\title{Observation of Power Superbroadening of Spectral Line Profiles on IBM Quantum}

\author{Ivo S. Mihov}
\affiliation{Center for Quantum Technologies, Department of Physics, Sofia University, 5 James Bourchier blvd, 1164 Sofia, Bulgaria}
\author{Nikolay V. Vitanov}
\affiliation{Center for Quantum Technologies, Department of Physics, Sofia University, 5 James Bourchier blvd, 1164 Sofia, Bulgaria}

\date{\today }

\begin{abstract}
Power broadening refers to the widening of the spectral line profile in a two-state quantum transition as the strength of the driving field increases. 
This phenomenon commonly arises in continuous-wave driving when the radiation field's intensity exceeds the transition's saturation intensity and it has been extensively studied in spectroscopy. 
For pulsed-field excitation, the spectral response of the quantum system may differ significantly:
while a rectangular-shaped pulse leads to a linear power broadening, pulses with smooth shapes show significantly reduced power broadening, for instance, logarithmic for the Gaussian shape and none for the hyperbolic-secant shape.
Recently [Phys. Rev. Lett. \textbf{132}, 020802 (2024)], in a dramatic paradigm shift, we have demonstrated experimentally that for Lorentzian-shaped pulses, the opposite effect --- power narrowing --- takes place: the width of the spectral profile decreases when the driving pulse amplitude increases, with a narrowing factor of as much as 10 observed.

While in high-resolution spectroscopy the push is for eliminating or even inverting the power broadening, there are applications where it is used to an advantage for it facilitates off-resonance excitation.
Here, we present a number of shaped pulses that exhibit power broadening much greater than that of the rectangular pulse of the same pulse area.
They are grouped in two families of pulse shapes.
In particular, in regard to the width of the second Rabi oscillation maximum, the quadratic pulse family shows a spectral increase by a factor of 3.3 whereas the power-law pulse family exhibits an increase by a factor of more than 3.5.

\end{abstract}

\maketitle


\section{Introduction}
%
%

Power broadening is a well-known phenomenon in spectroscopy, causing significant interaction between light and matter in a broader frequency spectrum away from resonance. 
While it is often associated with undesirable effects, such as reduced sensitivity in high-resolution spectroscopy and a cause for unwanted transitions (cross talk) in qubit control, there are applications where it is found to be beneficial. 
For instance, the same mechanism by which it activates unwanted transitions can be used to enable weak but desired transitions. 
One such application is laser cooling, where power broadening facilitates effective transitions even with large detunings ~\cite{Lindenfelser2017, Prudnikov2019, Fedorov2015, Mao2023}.
Additionally, power broadening enables the preparation of certain quantum states in cases where necessary transitions are not accessible otherwise~\cite{Park2011}.
Finally, power broadening is widely used in spectroscopy to effectively increase the width of narrow resonance peaks that are difficult to detect, enabling their capture and accurate measurement~\cite{Berry2006, Ya2001, Levine2009, Calderon2008}.
Power-broadened slowing of atoms and molecules plays a key role in robust loading of particles into magneto-optical traps~(MOTs)~\cite{Zhu1991, Yeo2015, Ravi2025, Williams2017, Truppe2017}.
By helping the interaction occur in a broader range of frequencies, the resonance requirement becomes less strict.
Therefore, interactions that are otherwise hindered by the resonance condition are enabled.

The power broadening effect has been a research area of interest in quantum physics and spectroscopy ever since early studies on the phenomenon~\cite{Breene1961, Breene1981, Breene19812, Allen1975, Shore1990, Citron1977, Bergquist1996}.
Theoretical interest in the effect has never ceased~\cite{Foot2005, Milonni2010, Vitanov2001, Halfmann2003, Harth1995, Roy2011, Levine2012, Bettles2020, Conover2011}.
A substantial body of research has focused on mitigating its effects to increase the precision of various spectroscopic techniques~\cite{Mihov2024, Mihov2023, Florez2013, Reimer2016}. 
Indeed, relaxing the resonance condition complicates the accurate determination of the natural frequency, which is fundamental to high-resolution spectroscopy.
The impact is generally negative, as it forces spectroscopists to limit the power of the pulses they use for finding the spectral line width, which results in very long data acquisition times, often many weeks and months~\cite{Dreissen2022}.

Off-resonant qubit excitations are affected by the temporal shape of the Rabi frequency envelope.
Hence, the pulse shape properties have long been used to solve common problems in quantum systems, such as leakage, crosstalk, noise sensitivity, etc.~\cite{Mihov2023, Mihov2024, Mihov20242, Motzoi2009, Li2024, Zhu2022, Harutyunyan2023, Peyraut2023, Kuzmanovic2024, Mccord2025, Boradjiev2013pra, Saharyan2023, Dridi2024}.
As a promising technique for exerting control on the lineshape, pulse shaping allows to reduce or completely eliminate the power broadening of the qubit spectral line.
Power refers to the drive power at the qubit input. 
For fixed wiring and calibration, $P\propto \Omega_0^2$, so we report and control power via the Rabi frequency $\Omega_0/2\pi$.
Gaussian, Sine, Sine$^2$ and other temporal pulse envelopes can be used to diminish the broadening effect, while the sech pulse halts it completely~\cite{Rosen1932, Vitanov2001, Mihov2023}.
Some earlier works even study the potential of pulse shape families that reverse the power broadening \cite{Robinson1985, Boradjiev2013, Mihov2024}, introducing the concept of power narrowing using Lorentzian and Lorentzian-based temporally-shaped pulses.

In this work, we investigate how to achieve the opposite effect -- generation of maximal power broadening. 
Starting from the most common example of a linear-broadening-generating rectangular pulse shape, we aim to increase the non-adiabaticity of the pulse envelope, a metric that is shown in Sec.~\ref{sec-pulseshapes} to predict the degree of power broadening caused by the utilized pulse.
We achieve stronger power broadening than the one produced by rectangular pulse shapes with the help of two families of pulse shapes with inherent highly non-adiabatic artefacts.

\section{Adiabatic Evolution in Two-State Systems}

\subsection{Two-State Schr\"odinger Equation}

In a two-state quantum system, the time-dependent Schr\"odinger equation governs the evolution of the state vector $\mathbf{C}(t)$ by the equation (in natural units $\hbar=1$)
\begin{equation}
    i \frac{d}{dt}  \mathbf{C}(t) = \mathbf{H}(t) \mathbf{C}(t),
\label{eq-schro}
\end{equation}
where \( \mathbf{H}(t) \) is the time-dependent Hamiltonian of the system and $\mathbf{C}(t) = \left[c_1(t), c_2(t)\right]^T$ is the state vector, consisting of $c_1(t)$ and $c_2(t)$ -- the (complex) probability amplitudes of the two states.

\subsection{Hamiltonian in Two-State System}
The Hamiltonian for a two-state system in the rotating-wave approximation (RWA) can be represented as
\begin{equation}
    \mathbf{H}(t) = \frac{1}{2} \begin{bmatrix} -\Delta(t) & \Omega(t) \\ \Omega^*(t) & \Delta(t) \end{bmatrix},
\end{equation}
where \( \Delta(t) \) is the detuning and \( \Omega(t) \) is the Rabi frequency. 
Its eigenvalues are
\be
\varepsilon_{\pm}(t) = \pm \frac{1}{2} \sqrt{\Omega^2(t) + \Delta^2(t)}.
\label{eq-eigenvals}
\ee

In the context of pulse shaping, both \( \Delta(t) \) and \(\Omega(t)\) can vary with time.
Substituting this Hamiltonian into Eq.~\eqref{eq-schro}, we obtain the two-state Schr\"odinger system of equations 
\begin{align}
    \nonumber
    i \dot{c}_1(t) &= -\frac12\Delta(t)c_1(t) + \frac12\Omega(t) c_2(t),  \\
    i \dot{c}_2(t) &= \frac12\Omega^*(t)c_1(t) + \frac12\Delta(t) c_2(t),
\label{eq-schro-2level}
\end{align}
where the dot over the probability amplitudes $c_1(t)$ and $c_2(t)$ represents the derivative of the corresponding variable with respect to time.

\subsection{Adiabatic Evolution}\label{sec-adiabevol}
Adiabatic evolution occurs when the system’s Hamiltonian changes slowly enough over time so the system remains in an instantaneous eigenstate of $\mathbf{H}(t)$, provided it started in one. This implies that the system avoids transitions between eigenstates, evolving within the same eigenstate~\cite{Born1928}.
This abstinence from transitioning only occurs in the adiabatic basis, while in the bare basis, transitions between the ground and excited states may occur only in the presence of chirp, change in detuning, $\dot{\Delta}(t)$.

\subsection{Boosting power broadening -- non-adiabaticity}

The adiabatic condition requires the change in the Hamiltonian to be slow relative to the energy difference between the eigenstates.
If this is satisfied, the system evolves adiabatically.
In quantum power broadening and power superbroadening --- an enhanced form of power broadening which exceeds the broadening manifested by the rectangular pulse shape ---  pulse shaping affects both the Rabi frequency \( \Omega(t) \) and the adiabaticity conditions, making it possible to maintain control over transitions and achieve substantial broadening effects.
This connection has been studied in relation to power narrowing in Ref.~\cite{Mihov2024}.
Expanding on the ideas in that study, we regard power broadening as a consequence of the nonadiabatic nature of the pulse shape.
Therefore, we demand significant non-adiabaticity to achieve strong power broadening.
This is the reason we look for pulse shapes that have more non-adiabatic artifacts than the rectangular pulse and naturally violate the adiabatic condition
\begin{equation}
\underbrace{\sqrt{\Omega^2(t) + \Delta^2(t)}}_{\varepsilon_1(t)} \gg \underbrace{\frac{\dot{\Omega}(t)\Delta(t) + \Omega(t)\dot{\Delta}(t)}{\Omega^2(t) + \Delta^2(t)}}_{\dot{\vartheta_1}(t)},
\label{eq-adiabcond}
\end{equation}
where $\varepsilon_1(t)$ is the adiabatic splitting and $\dot\vartheta_1(t)$ is the nonadiabatic coupling.
Whenever $\varepsilon_1(t)$ is larger than $\dot\vartheta_1(t)$ we have adiabatic evolution, described in Subsec.~\ref{sec-adiabevol}.
However, if $\varepsilon_1(t)$ is smaller than $\dot\vartheta_1(t)$, we see nonadiabatic effects such as highly off-resonant transitions and power broadening.
Thus, we seek nonadiabatic pulse shapes that exhibit sudden Rabi frequency changes concurrent with small Rabi frequencies.

The theory of adiabaticity begins with the construction of the adiabatic states --- the (time-dependent) eigenstates of the Hamiltonian~\cite{Boradjiev2013pra, Born1928}. 
The original diabatic basis is related to the adiabatic one by the transformation matrix $\mathbf{R}(\vartheta_1)$: 
\be
\mathbf{C}(t) = \mathbf{R}(\vartheta_1(t)) \mathbf{A}_1(t),
\ee
where
\be
\mathbf{R}(\vartheta_1) = 
\begin{bmatrix}
\cos \vartheta_1 & \sin \vartheta_1 \\
-\sin \vartheta_1 & \cos \vartheta_1
\end{bmatrix},
\ee
with
\be
\vartheta_1(t) = \frac{1}{2} \arctan \left(\frac{\Omega(t)}{\Delta}\right) \quad (0 \leq \vartheta_1(t) \leq \pi/4).
\label{eq-mixingangle}
\ee

Substituting the diabatic states $\mathbf{C}(t)$ into the Schr\"odinger equation~\eqref{eq-schro}, we rearrange and obtain an analogical equation, extracting the adiabatic Hamiltonian 

\begin{align}    
\mathbf{H}_1(t) &= \mathbf{R}(-\vartheta_1) \mathbf{H} \mathbf{R}(\vartheta_1) 
- i \mathbf{R}(-\vartheta_1) \frac{d}{dt} \mathbf{R}(\vartheta_1)\\
&=\left(\begin{array}{cc}
    \varepsilon_-(t) & -i\dot{\vartheta_1}(t) \\
    i\dot{\vartheta_1}(t) & \varepsilon_+(t)
\end{array}\right),
\label{eq-adb-ham}
\end{align}
where $\varepsilon_\pm$ are the eigenvalues of the diabatic Hamiltonian from Eq.~\eqref{eq-eigenvals}.
Now we have a Hamiltonian which becomes diagonal (i.e. we stay in the adiabatic states) if $\varepsilon_+-\varepsilon_- \gg \left|\dot{\vartheta_1}\right|$. 
Since $\varepsilon_+(t)-\varepsilon_-(t) = \varepsilon_1(t) = \sqrt{\O^2(t)+\Delta^2}$, we obtain the adiabatic condition, shown in Eq.~\eqref{eq-adiabcond}.


\begin{figure*}[ht!]
    \centering
    \includegraphics[width=1\linewidth]{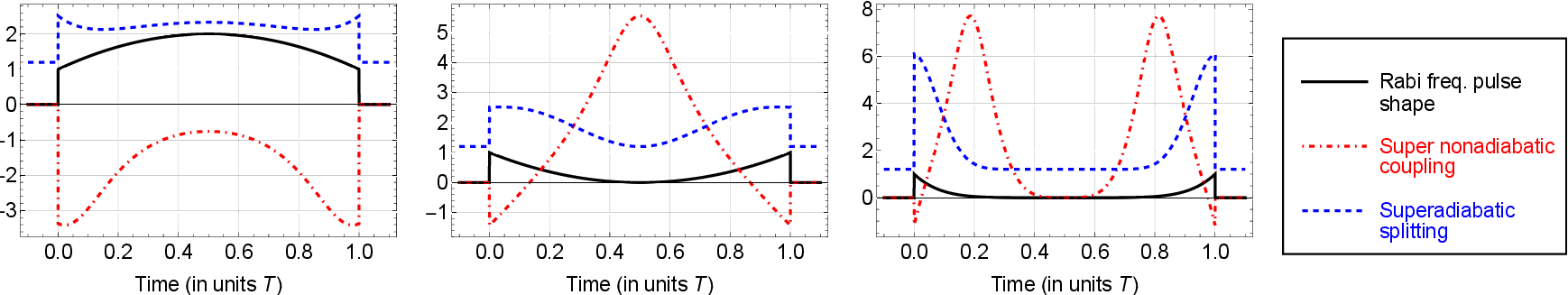}
    \caption{[Color online] Super-nonadiabatic coupling (dot-dashed curves) and superadiabatic energy splitting (dashed curves) for three drive envelopes (solid curves), shown left-to-right: 
    (i) quadratic pulse with $\beta = –1$ (see Eq.~\eqref{eq-fcquadratic}),
    (ii) quadratic pulse with $\beta = +1$ (see Eq.~\eqref{eq-fcquadratic}), and 
    (iii) power-law pulse with $P = 3$ (see Eq.~\eqref{eq-evenexp}).
    All traces share the same time axis and peak Rabi amplitude, allowing a direct comparison of how the sign of $\beta$ and the power-law exponent $P$ modify the nonadiabatic region near the pulse center.
    }
    \label{fig:superadb}
\end{figure*}


\subsection{Superadiabatic basis}

The superadiabatic (SA) bases~\cite{Boradjiev2013pra, Berry1990, Joye1993, Drese1998} can be used as a diagnostic to identify and quantify nonadiabatic regions generated by our pulse shape families.
The (super)adiabatic bases may facilitate easier simulation when the evolution is (super)adiabatic and the corresponding condition is satisfied.
In contrast, should one need to simulate the qubit dynamics when the evolution is nonadiabatic, it would be instead advantageous to perform exact numerical solution of the two-state Schr\"odinger equation, rather than use (super)adiabatic evolution.
The superadiabatic bases are obtained by repeated diagonalization of the previous-order superadiabatic bases.
We regard the adiabatic basis an order-1 superadiabatic basis, and use it as a starting point in order to retrieve the $n=2$ basis.
We begin with a phase transformation on the Hamiltonian,
\be
\mathbf{F} = 
\begin{bmatrix}
e^{-i\pi/4} & 0 \\
0 & e^{i\pi/4}
\end{bmatrix},
\label{eq-phase_transform}
\ee

Thus, we get the transformation $\mathbf{H}_1(t) \rightarrow \mathbf{H}_2(t) = \mathbf{F}^\dagger \mathbf{H}_1(t) \mathbf{F}$, as well as some analogues of the $n=1$ variables,

\be
\begin{aligned}
\Delta &\rightarrow \varepsilon_1(t),\\
\Omega(t) &\rightarrow 2\dot{\vartheta}_1(t),\\
\mathbf{C}(t) &\rightarrow \mathbf{A}_1(t).
\end{aligned}
\ee

Indeed, all of the adiabatic variables have their analogues,
\be
\begin{aligned}
\vartheta_1(t) &\rightarrow \vartheta_2(t), \\
\varepsilon_1(t) &\rightarrow \varepsilon_2(t), \\
\mathbf{A}_1(t) &\rightarrow \mathbf{A}_2(t), \\
\mathbf{H}_1(t) &\rightarrow \mathbf{H}_2(t), \\
\eta_1 &\rightarrow \eta_2.
\end{aligned}
\ee
The superadiabatic condition becomes 
\be
\underbrace{\sqrt{4\dot{\vartheta}_1^2(t) + \varepsilon_1^2(t)}}_{\varepsilon_2(t)} \gg \underbrace{\frac{2\ddot{\vartheta}_1(t) \varepsilon_1(t) + 2\dot{\vartheta}_1(t) \dot{\varepsilon}_1(t)}{4\dot{\vartheta}_1^2(t) + \varepsilon_1^2(t)}}_{\dot{\vartheta}_2(t)}.
\label{eq-superadb-cond}
\ee
By repeating the process, we can also obtain the higher-order superadiabatic bases~\cite{Boradjiev2013pra, Berry1990, Joye1993, Drese1998}.

The super-nonadiabatic coupling and the SA splitting are plotted for three pulse shapes in Fig.~\ref{fig:superadb} to help find nonadiabatic spots that predict the resulting power superbroadening.

\subsection{Relation between adiabatic and superadiabatic conditions}
\label{sec:SA-condition}

Using the usual mixing angle from Eq.~\eqref{eq-mixingangle}, its time derivative, the nonadiabatic coupling from Eq.~\eqref{eq-adiabcond}, reads
\begin{equation}
\dot{\vartheta}_1(t)=\frac{\dot{\Omega}(t)\,\Delta(t)-\Omega(t)\,\dot{\Delta}(t)}{2\,[\Omega^2(t)+\Delta^2(t)]}.
\label{eq:thetadot1}
\end{equation}
The adiabatic condition \eqref{eq-adiabcond} is $\varepsilon_1\gg|\dot{\vartheta}_1|.$
After one SA transformation, the splitting and the coupling are shown in Eq.~\eqref{eq-superadb-cond}:
\begin{align}
&\varepsilon_2(t)=\sqrt{\varepsilon_1^2(t)+4\,\dot{\vartheta}_1^2(t)},\\
&\dot{\vartheta}_2(t)=\frac{2\,\ddot{\vartheta}_1(t)\,\varepsilon_1(t)+2\,\dot{\vartheta}_1(t)\,\dot{\varepsilon}_1(t)}
{\varepsilon_1^2(t)+4\,\dot{\vartheta}_1^2(t)}.
\label{eq:theta2eps2}
\end{align}

For pulses characterized by a timescale $\tau$, the smoothness implies
$|\dot{\Omega}|,|\dot{\Delta}|=\mathcal{O}(\tau^{-1})$ and
$|\ddot{\Omega}|,|\ddot{\Delta}|=\mathcal{O}(\tau^{-2})$. 
Hence
\begin{equation}
|\dot{\vartheta}_1|=\mathcal{O}\!\big((\varepsilon_1\tau)^{-1}\big),\qquad
|\dot{\vartheta}_2|=\mathcal{O}\!\big(\varepsilon_1^{-3} \tau^{-2}\big).
\end{equation}
Since $\varepsilon_2\ge \varepsilon_1$, if \eqref{eq-adiabcond} holds with margin
$\gamma=\varepsilon_1/|\dot{\vartheta}_1|\gg 1$, then
\begin{equation}
\frac{\varepsilon_2}{|\dot{\vartheta}_2|}\;\gtrsim\;\mathcal{O}(\gamma^2),
\end{equation}
so the SA condition \eqref{eq-superadb-cond} is not stricter than \eqref{eq-adiabcond}; it is typically easier to satisfy.
Conversely, violating \eqref{eq-superadb-cond} marks a strongly nonadiabatic (“super-nonadiabatic”) regime where the SA coupling is large.

These observations motivate our envelope design principle: maximize $|\dot{\Omega}(t)|$ where $|\Omega(t)|$ is small (mid-pulse pits or sharp curvature). 
This maximizes $|\dot{\vartheta}_1|$ and $|\dot{\vartheta}_2|$ near the pulse center, producing the observed power superbroadening.


This asymptotic implication ``adiabatic $\Rightarrow$ superadiabatic'' refers to the regime where \eqref{eq-adiabcond} holds with a large margin; it does not assert a pointwise ordering when the adiabatic condition is only marginally satisfied or violated.

For $\dot\Delta=0$,
\be
\dot{\vartheta}_1=\frac{\Delta\,\dot{\Omega}}{2(\Omega^2+\Delta^2)}
=\frac{\Delta\,\dot{\Omega}}{2\varepsilon_1^2},\qquad
\varepsilon_1=\sqrt{\Omega^2+\Delta^2}.
\ee
Hence the obstacle to adiabatic evolution depends essentially on the slope $|\dot{\Omega}|$.
By contrast, the second--order coupling reads
\be
\dot{\vartheta}_2
=\frac{2\,\ddot{\vartheta}_1\,\varepsilon_1+2\,\dot{\vartheta}_1\,\dot{\varepsilon}_1}
{\varepsilon_1^2+4\dot{\vartheta}_1^{\,2}}
=\frac{\Delta\big(\varepsilon_1^2\,\ddot{\Omega}-\Omega\,\dot{\Omega}^{\,2}\big)}
{\varepsilon_1^3\big(\varepsilon_1^2+4\dot{\vartheta}_1^{\,2}\big)},
\ee
where $\dot{\varepsilon}_1=\Omega\,\dot{\Omega}/\varepsilon_1$ and
$\ddot{\vartheta}_1=\Delta(\varepsilon_1^2\,\ddot{\Omega}-2\,\Omega\,\dot{\Omega}^{\,2})/(2\,\varepsilon_1^4)$.
Near the pulse center of the chirp-free quadratic family, where $|\Omega|\ll|\Delta|$ so that
$\varepsilon_1\simeq|\Delta|$, one obtains the form
\be
\dot{\vartheta}_2 \;\simeq\;
\mathrm{sgn}(\Delta)\;
\frac{\,\ddot{\Omega}\;-\;(\Omega/|\Delta|^2)\,\dot{\Omega}^{\,2}\,}
{\,|\Delta|^2\Big(1+\frac{\dot{\Omega}^{\,2}}{|\Delta|^{4}}\Big)}\,.
\ee
Thus the $4\dot{\vartheta}_1^{\,2}$ contribution is retained through the positive regularizing factor $1/(1+\dot{\Omega}^{\,2}/|\Delta|^{4})$.
Using the adiabatic margin $\gamma_1=\varepsilon_1/|\dot{\vartheta}_1|\simeq 2|\Delta|^{2}/|\dot{\Omega}|$, we have
\be
\frac{\dot{\Omega}^{\,2}}{|\Delta|^{4}}=\frac{4}{\gamma_1^{2}},
\ee
and therefore
\be
\dot{\vartheta}_2 \;\simeq\;
\mathrm{sgn}(\Delta)\;
\frac{\,\ddot{\Omega}\;-\;(\Omega/|\Delta|^2)\,\dot{\Omega}^{\,2}\,}
{\,|\Delta|^2\,}\;\cdot\;\frac{\gamma_1^{2}}{\gamma_1^{2}+4}\,.
\ee
Two regimes follow immediately:\\
(i) if $\gamma_1\gg 1$ and $|\Omega|\ll|\Delta|$, the denominator factor $\gamma_1^{2}/(\gamma_1^{2}+4)\to 1$ and $\dot{\vartheta}_2\simeq \mathrm{sgn}(\Delta)\,\ddot{\Omega}/|\Delta|^{2}$ (curvature–controlled);\\
(ii) if $\gamma_1=\mathcal{O}(1)$, the $4\dot{\vartheta}_1^{\,2}$ term cannot be neglected and it suppresses $\dot{\vartheta}_2$ by the factor $\gamma_1^{2}/(\gamma_1^{2}+4)$.\\
In both cases the denominator is positive, so the sign of $\dot{\vartheta}_2$ (and the $\beta=\pm 1$ contrast) is governed by the curvature term $\ddot{\Omega}$ instead of the slope $\dot\Omega$, while the denominator only rescales its magnitude.

Consequently, two chirp-free envelopes that mildly violate (or similarly satisfy) the adiabatic condition --- because they have comparable $|\dot{\Omega}|$ --- can still produce markedly different dynamics at second order due to opposite curvature near the midpoint.
This explains the observed disparity between $\beta=-1$ and $\beta=+1$ and precisely shows the diagnostic value of the SA basis in our work.
Furthermore, we can see in Ref.~\cite{Boradjiev2013pra} how higher-order SA violations can be used to map the behavior of higher-exponent sinusoidal pulse shapes.


\begin{figure}
    \centering
    \begin{subfigure}[tbph]{0.48\textwidth}
        \centering
        \includegraphics[width=\textwidth, height=1.25\textwidth]{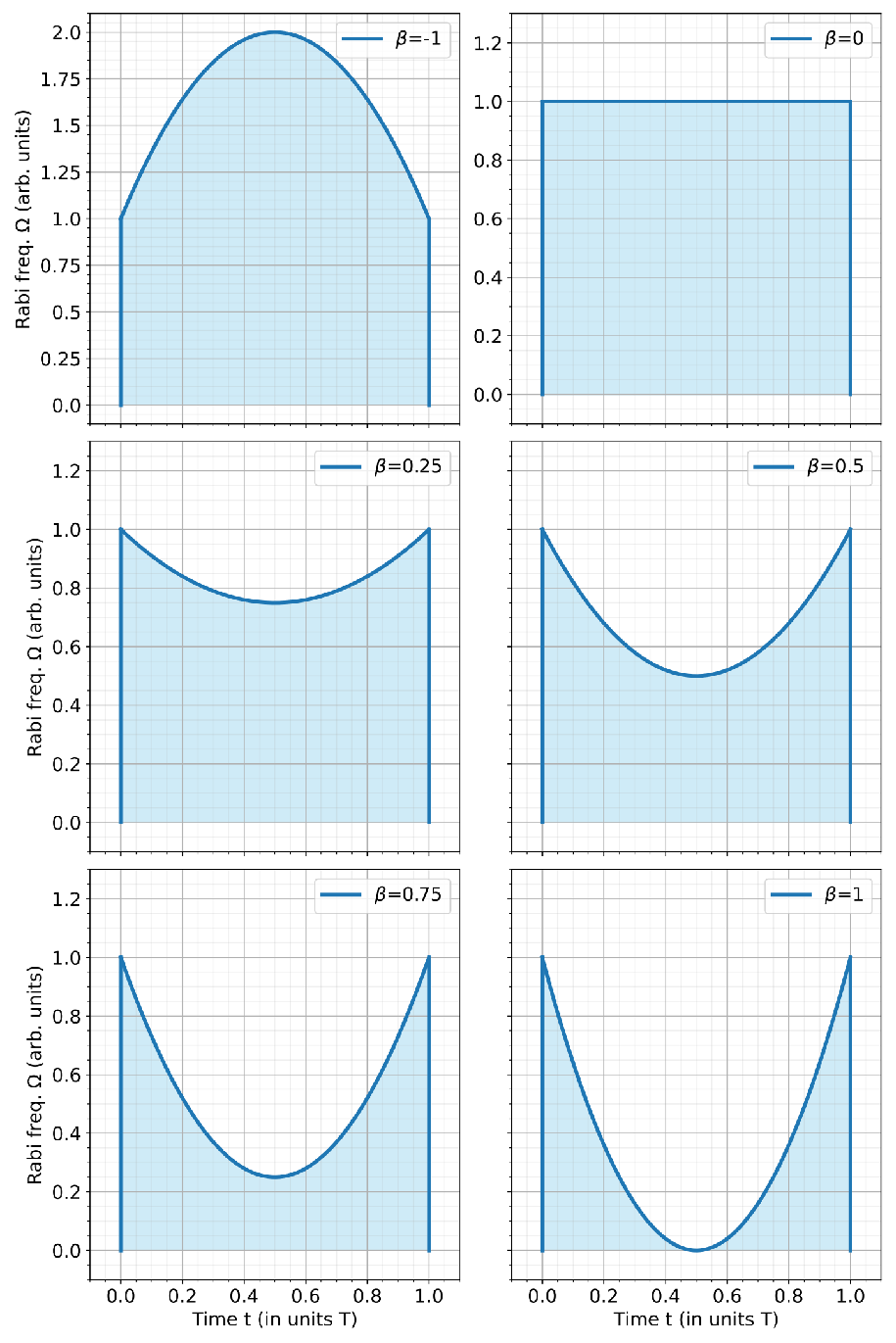}
        \caption{}
        \vspace{10pt}
        \label{fig-fcq-pulse-shapes}
    \end{subfigure}
    \hfill
    \begin{subfigure}[tbph]{0.48\textwidth}
        \centering
        \includegraphics[width=\textwidth]{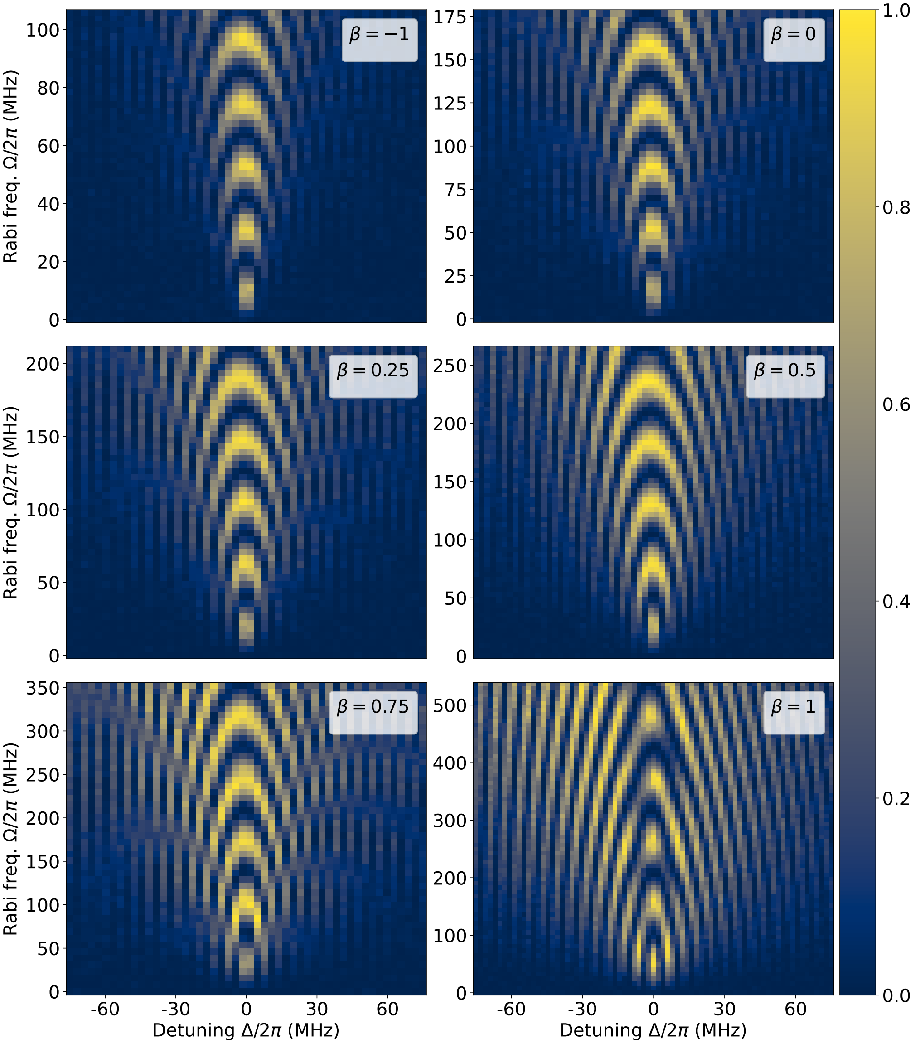}
        \caption{}
        \label{fig-fcq-excitation-landscapes}
    \end{subfigure}
    \label{fig:fcq}
    \caption{[Color online] \textbf{Top:} Quadratic pulse envelopes $\Omega(t)$ for $\beta\in\{-1,0,0.25,0.5,0.75,1\}$ at fixed duration.
    \textbf{Bottom:} Final excited-state population $P_2(T)$ versus detuning $\Delta/2\pi$ (x-axis) and Rabi amplitude $\Omega_0/2\pi$ (y-axis, $\propto \sqrt{\text{power}}$).
    The operational linewidth is defined as the full excitation width in $\Delta$ of the second Rabi maximum.
    Increasing $\beta$ enhances nonadiabatic coupling near the pulse midpoint (see Fig.~\ref{fig:superadb}), yielding pronounced power superbroadening relative to the rectangular case ($\beta=0$).
    }
\end{figure}

\begin{figure}
    \centering
    \begin{subfigure}[tbph]{0.48\textwidth}
        \centering
        \includegraphics[width=\textwidth, height=0.95\textwidth]{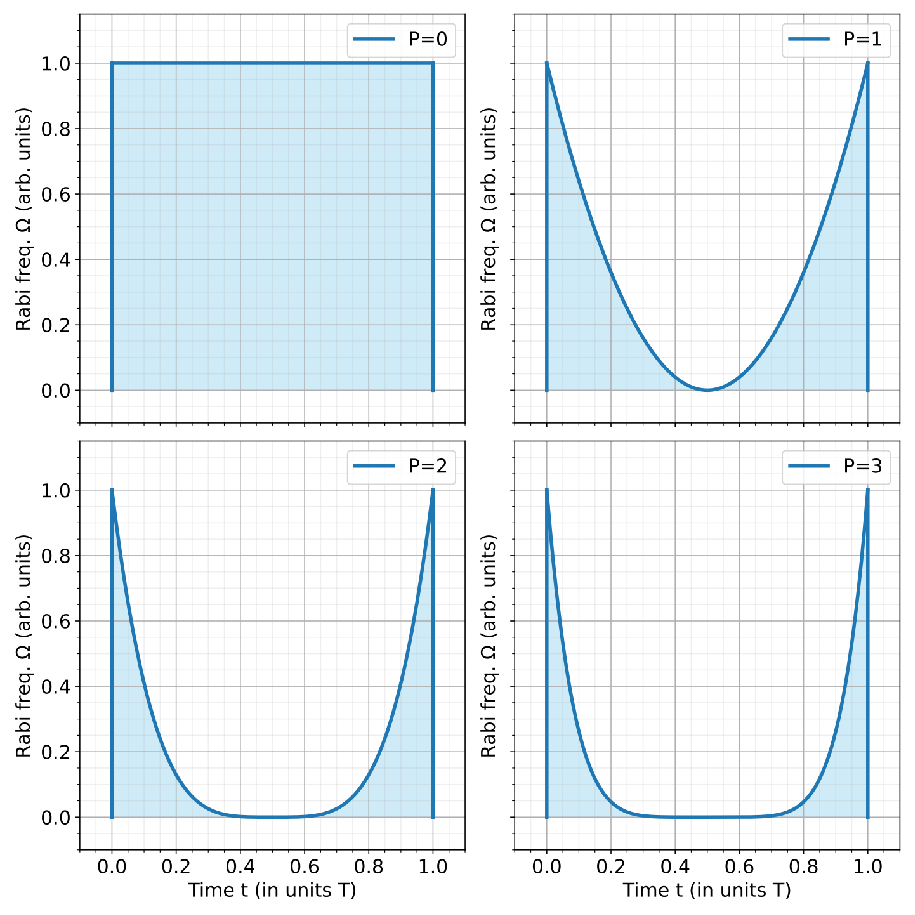}
        \caption{}
        \vspace{10pt}
        \label{fig-ee-pulse-shapes}

    \end{subfigure}
    \hfill
    \begin{subfigure}[tbph]{0.48\textwidth}
        \centering
        \includegraphics[width=\textwidth]{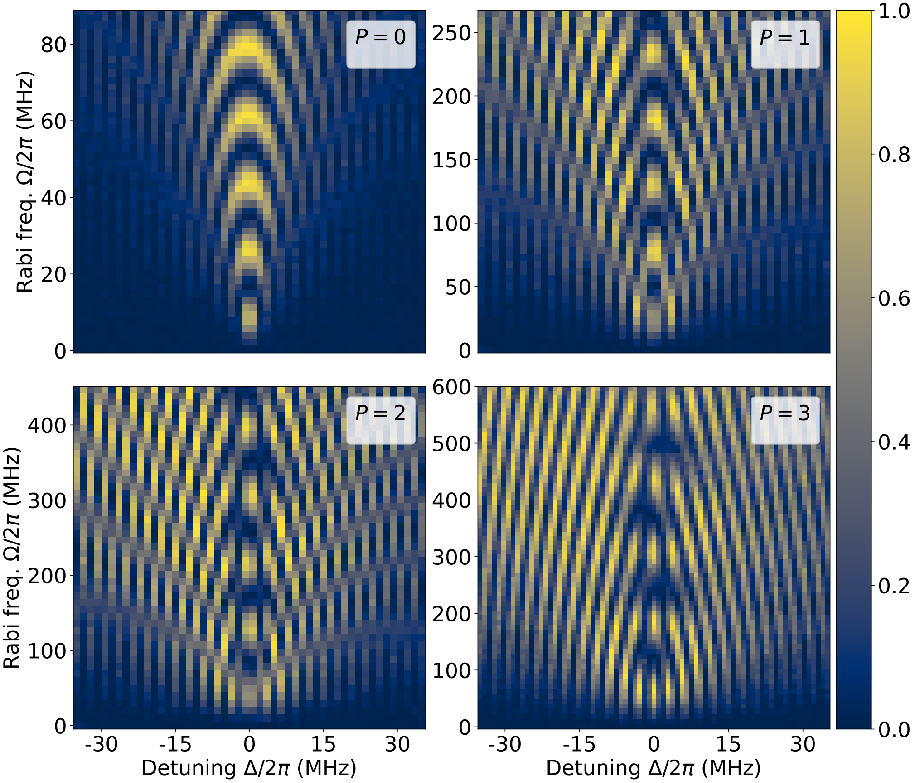}
        \caption{}
        \label{fig-ee-excitation-landscapes}

    \end{subfigure}
    \label{fig:ee}
    \caption{[Color online] \textbf{Top:} Power-law pulse envelopes $\Omega(t)$ with a mid-pulse pit of tunable width for $P=0,1,2,3$ (rectangular pulse at $P=0$).
    \textbf{Bottom:}  Excitation landscapes. 
    For $P>0$ the two horns act as Ramsey-like $\pi/2$ drives separated by a dark interval, yielding near-vertical interference fringes off-resonance. 
    The pronounced increase of the operational linewidth is an indicator of power superbroadening for growing area.
    }
\end{figure}

\section{The quadratic and the power-law pulse shapes}
\label{sec-pulseshapes}

We seek envelopes that produce large $|\dot{\vartheta}_1|$ (and, at second order, $|\dot{\vartheta}_2|$) at small $|\Omega|$, thereby enhancing nonadiabatic coupling away from resonance.
The quadratic family flips curvature mid-pulse, isolating the role of slope sign ($\beta=-1$ vs. $\beta=+1$), while the power-law family enforces a mid-pulse pit of tunable width (parameter $P$), effectively creating two separated horns reminiscent of Ramsey $\pi/2$–$\pi/2$ sequences. Both families \\
(i) keep $\Omega(t)\ge 0$ (no phase flips), \\
(ii) have a characteristic timescale $\tau$ given by $T/2$, \\
(iii) control non-adiabaticity with a single parameter, \\
(iv) include the rectangular pulse as a limit, and \\
(v) are easily applied under Qiskit amplitude/duration constraints, enabling a clean hardware demonstration.
We then proceed to demonstrate the models' excitation landscapes --- post-pulse excited population versus detuning (x-axis) and Rabi frequency (y-axis) --- on IBM's \textsc{ibm\_sherbrooke} (see specifics of the demonstration in Appendix~\ref{sec-params}).

The quadratic pulse shape can be regarded as a rectangular pulse with an added parabolic term, as illustrated in Fig.~\ref{fig-fcq-pulse-shapes}. 
Despite the simplicity of its crude form -- $\Omega(t) \propto 1 + \beta t^2$, where $\beta$ is a parameter -- the normalized expression for this pulse shape is 
\be
\Omega(t) = \Omega_0 \left\{1 + \beta \left[\left(\frac{t-T/2}{T/2}\right)^2-1\right]\right\},
\label{eq-fcquadratic}
\ee
where $\beta$ controls the degree of concavity or convexity of the added parabola, and $T$ is the pulse duration.
The parameter $\beta$ has an applicability range $\beta \in (-\infty, 1]$. 
The upper boundary $\beta=1$ represents the case where the midpoint of the pulse touches the $\Omega(t)=0$ line.
For higher values of $\beta$, its Rabi frequency would become negative at the midpoint, an undesirable characteristic, which motivated us to introduce the upper limit of $\beta$.

Similarly designed to break the superadiabatic inequality, the power-law pulse shape incorporates nonadiabatic features, namely discontinuous jumps at the edges and an inherent zero value at the midpoint of the pulse.
Its crude form is $\Omega(t) \propto t^{2P}$, but adjusted for unity it becomes 
\be
\Omega(t) = \Omega_0 \left(\frac{t-T/2}{T/2}\right)^{2P},
\label{eq-evenexp}
\ee 
where $P\in \mathbb{N}$ is a parameter, \textit{half the exponent of time $t$}, and controls the width of the pit.
With $P=0$, the power-law pulse shape becomes the standard rectangular pulse shape, or put alternatively -- the pit has zero width. 
For $P \gg 1$ the shape approximates two Dirac delta functions time $T$ apart.
For a high power $P$ and the right pulse area, where each of the two horns has an area $\mathcal{A} = \pi/2$, combining for a total area of $\pi$, the pulse can be used to produce Ramsey fringes in the frequency spectrum.
Similar to the $\beta=1$ case of the quadratic pulse shape, the power-law pulse also inherently reaches zero in the midpoint for all values of its parameter $P$.

Interestingly, the $\beta=-1$ and $\beta=+1$ orientation-changing quadratic pulse shapes have the same derivatives in magnitude, only the signs are inverted.
Intuitively, this could incline us to believe that their representations in the adiabatic basis would not differ and that the adiabatic inequality would lead to similar conclusions for both of them.
However, one can see in Fig.~\ref{fig:superadb} that the superadiabatic inequality is violated by the $\beta=1$ quadratic pulse, and by the $P=3$ power-law pulse shape but not by the $\beta=-1$ quadratic pulse shape.
This finally allows us to see why the $\beta=-1$ and $\beta=+1$ shapes behave differently at $\Delta \neq 0$.

\subsection{Orientation-changing quadratic}

The orientation-changing quadratic pulse shapes with $\beta=-1,0,0.25,0.5,0.75 \text{, and } 1$ were tested on the 127-qubit IBM quantum computer --- \textsc{ibm\_sherbrooke}.
The detailed list of parameters recorded during the experiment is shown in Appendix~\ref{sec-params}.
The duration of all quadratic pulses was 800 \textit{dt} units --- the system time units ($dt=\frac29$ ns) used in IBM's Qiskit framework --- which corresponds to 177.8~ns.
As stated previously, the $\beta=0$ case of the orientation-changing quadratic is in fact the constant pulse. 
It arises from simply turning the pulse generator on and off after time $T$ and is known for reproducing the common linear power broadening.
In line with this hypothesis, we indeed see the common broadening pattern in the top right excitation map of Fig.~\ref{fig-fcq-excitation-landscapes}, corresponding to the $\beta=0$ case.
It is also evident that as $\beta$ changes from -1 to 0, the power broadening pattern grows stronger.
In the $\beta=-1$ plot the second maximum is about $20$~MHz wide, whereas in the $\beta=0$ case, a fringe appears at around $\Delta=\pm15$~MHz, resulting in a linewidth of 30~MHz.
The broadening pattern is much stronger in the linewidth of the $\beta=1$ pulse, spanning a frequency interval of about 100~MHz, or an increase by a factor of about 3.3 compared to the rectangular pulse ($\beta=0$).
It is not intuitive that the $\beta = -1$ case, albeit having the same derivative in magnitude, shows strikingly weaker power broadening than the $\beta=1$ case. 
These two examples form the limiting cases in the experiment, with the $\beta=1$ plot having at least 5 visible fringes around the second maximum.
Both pulse shapes have exactly the same derivatives in magnitude, but their signs are different.
In the $\beta=-1$ case, the derivative is first positive, then negative for the second half of the pulse, while in the $\beta=1$ case, they are in the opposite order.
This difference may seem insignificant at first, but completely changes the pulse's appearance in the adiabatic and superadiabatic pictures. 

The reason is that the super-nonadiabatic effects in the latter case are stronger around the midpoint of the pulse.
This is shown using the superadiabatic basis and its super-nonadiabatic coupling and super-splitting as evident from the comparison in Fig.~\ref{fig:superadb}. 
In the latter case $\O(t)$ is negligible in the midpoint area, while the derivatives $\dot{\O}(t)$ are the same.
In the middle plot of Fig.~\ref{fig:superadb}, it is clear that the nonadiabatic coupling is higher for the $\beta=1$ pulse, which enables the transition even far from resonance.

\subsection{Power-law pulse}

The excitation landscapes in Fig.~\ref{fig-ee-excitation-landscapes} demonstrate progressively stronger power broadening, accompanied by increasingly pronounced fringe patterns.
Aiming at higher non-adiabaticity, we evaluated power-law pulse shapes with $P\in \{0, 1, 2, 3\}$ on the 127-qubit IBM quantum computer -- \textsc{ibm\_sherbrooke}. 
The values of key parameters recorded throughout the demonstration are listed in Appendix~\ref{sec-params}.
The duration of all power-law pulses was 1600 \textit{dt} units, or 355.6~ns.
Notably, for all $P>0$, the Rabi frequency at the pulse midpoint is zero, a distinctive feature of the power-law pulse shape.
This characteristic enhances the pulse's nonadiabatic behavior, effectively mimicking a configuration resembling two well-separated pulses, each with an area of $\pi/2$, similar to those employed in Ramsey interferometry to generate Ramsey fringes.
Taking a look at the width of the spectral lines for a $\pi$-area pulse for $P=0$ and $P=3$, we see a huge broadening in the $P=3$ case.
A glance at the first maxima standing at $\mathcal{A}=\pi$ signals that it broadens from roughly $8$~MHz wide in the $P=0$ case to about $30$~MHz in the $P=3$ case.
While the linewidth of the $3\pi$ pulse is about $20$~MHz in the $P=0$ case, it is wider than the $70$~MHz scope of the plot depicting the $P=3$ case, obtained by taking all visible fringes into account.

Examining the two plots on the bottom row --- corresponding to $P=2$ (left) and $3$ (right) --- we find that dark nearly vertical lines have started to appear $\approx 3$~MHz apart due to interference between the left and right horns of the power-law pulse.
These lines are to increase both in number and contrast with higher numbers of $P$.

\section{Conclusions\label{Sec-conclusion}}

In conclusion, we have presented viable and simple pulse shaping alternatives to the commonly used rectangular pulse shape.
Without significant shifts in duration and shape, we employ two new families of temporal pulse shapes to observe threefold broadening in the effective linewidth of the qubit, measured at the same, minimal pulse area of $\pi$. 
These pulse shapes are at the heart of the newly demonstrated power superbroadening phenomenon, a stronger variation of power broadening.
It can be used in cases where power broadening is beneficial, contributing to its positive effects.

Usually, these are cases where power broadening is used to excite off-resonant transitions or look for resonant frequencies in a broad interval.
Some interesting applications also include laser cooling, where off-resonant transitions supply less energy than the transition needs, thereby extracting energy from the system and decreasing its temperature.
The method described here builds upon the commonly employed power broadening effect to boost it further and promise to enhance its function in many applications.


\section{Acknowledgments} 

We gratefully acknowledge the Karoll Knowledge Foundation for providing financial support to I.S.M. during the preparation of this manuscript. 
Their contribution was invaluable, not only to this work but in encouraging the author’s ongoing scientific endeavors. 
This research is supported by the Bulgarian national plan for recovery and resilience, Contract No. BG-RRP-2.004-0008-C01 (SUMMIT), Project No. 3.1.4, and by the European Union’s Horizon Europe research and innovation program under Grant Agreement No. 101046968 (BRISQ). 
We acknowledge the use of IBM Quantum services and the supercomputing cluster PhysOn at Sofia University for this work. 

The views expressed are those of the authors, and do not reflect the official policy or position of IBM or the IBM Quantum team.


\appendix

\begin{table}[]
    \centering
    \begin{tabular}{|c|c|c|c|}\hline
        Day & $T_1$ (\textmu s) & $T_2$ (\textmu s) & Readout Error (\%) \\ \hline
        27 Nov & 393.66 & 596.88 & 0.66 \\ 
        29 Nov & 321.33 & 697.45 & 0.76 \\ 
        1 Dec & 316.42 & 492.27 & 0.58 \\ 
        3 Dec & 256.01 & 287.59 & 2.05 \\ 
        5 Dec & 347.80 & 468.87 & 1.58 \\ \hline
    \end{tabular}
    \caption{Hardware parameter values for qubit 46 of the \textsc{ibm\_sherbrooke} quantum system on 5 service days in 2024.}
    \label{tab-params}
\end{table}

\section{Quantum hardware demonstration specifics}
\label{sec-params}

The two pulse shape families were tested against measurements on qubit 46 of the 127-qubit transmon quantum processor -- IBM's \textsc{ibm\_sherbrooke}, an Eagle r3 Processor~\cite{IBM, Koch2007}.

Due to the limited free slots on the hardware, the measurements were taken on 5 different days.
The $P=0$ power-law pulse was taken on 27 Nov 2024, while the $P=1,2,3$ pulses were recorded on 29 Nov 2024.
The $\beta=-1$ and $0$ quadratic pulses were demonstrated on 1 Dec 2024, the $\beta=1$ pulse test was performed on 3 Dec 2024, and the $\beta = 0.75, 0.5$ and $0.25$ quadratic pulses --- on 5 Dec 2024.

Qubit 46 of the \textsc{ibm\_sherbrooke} system had a excitation frequency of 4.6741~GHz and an anharmonicity of -0.3134~GHz, both of which were constant over the days of the demonstration.
The $T_1$ decoherence times were between 250 and 400~\textmu s, while the $T_2$ were even better --- between 280 and 700~\textmu s.
A more detailed breakdown of the parameters on each day of the demonstration can be found in Table~\ref{tab-params}.
The recorded values imply a $T_2$-limited linewidth of several kHz, which is tiny compared to a linewidth between 5 and 150 MHz.
Thus, the dephasing effects in this demonstration are insignificant.
Each datapoint in the excitation landscapes of Figs.~\ref{fig-fcq-excitation-landscapes} and \ref{fig-ee-excitation-landscapes} is an average of 200 shots.

We use the Qiskit Pulse framework for Python to access the low-level settings of the quantum processor, reproducing the target pulse shapes.
Albeit freely available, the framework has some limitations due to its characteristics: \\
(i) a soft upper limit on the total duration of the pulse (usually at several \textmu s); \\
(ii) an upper limit on the amplitude (limited to 1 in Qiskit's arbitrary units); \\
(iii) a lower limit on the pulse duration caused by the discretization -- $2/9$~ns; \\
(iv) basis measurements of the first two energy levels only, i.e. all the leakage is summed together with the excited state population. \\

The framework accommodates a wide choice of resources that can be used to construct gates, methods, and algorithms.
Using some neat workarounds, the aforementioned limitations may not pose a problem, rather an immediate impediment.


\begin{thebibliography}{99}




\bibitem{Lindenfelser2017} F. Lindenfelser, M. Marinelli, V. Negnevitsky, S. Ragg and J. P. Home, New J. Phys. \textbf{19}, 063041 (2017).
 
\bibitem{Prudnikov2019} O. N. Prudnikov, R. Ya. Il'enkov, A. V. Taichenachev, and V. I. Yudin, Phys. Rev. A \textbf{99}, 023427 (2019).

\bibitem{Fedorov2015} S. A. Fedorov, G. A. Vishnyakova, E. S. Kalganova, D. D. Sukachev, A. A. Golovizin, D. O. Tregubov, K. Yu. Khabarova, A. V. Akimov, N. N. Kolachevsky, and V. N. Sorokin, Appl. Phys. B \textbf{121}, 275 (2015).

\bibitem{Mao2023} Y. Mao, J. Xu, S. Guan, H. Ji, W. Liu, D. Chen, Q. Gong, Y. Quan, X. Long, H. Luo, Z. Tan, arXiv:2312.09635 (2023).


\bibitem{Park2011} H. Park, E. S. Shuman, and T. F. Gallagher, Phys. Rev. A \textbf{84}, 052708 (2011).


\bibitem{Berry2006} J. J. Berry, M. J. Stevens, R. P. Mirin, and K. L. Silverman, Appl. Phys. Lett. \textbf{88}, 061114 (2006).

\bibitem{Ya2001} S. Ya. Tochitsky, C. Filip, R. Narang, C. E. Clayton, K. A. Marsh, and C. Joshi, Opt. Lett. \textbf{26}, 813 (2001).

\bibitem{Levine2009} J. Levine, M. R. Savina, T. Stephan, N. Dauphas, A. M. Davis, K. B. Knight, and M. J. Pellin, Int. J. Mass. Spectr. \textbf{288}, 36 (2009).

\bibitem{Calderon2008} O. G. Calderón, S. Melle, M. A. Antón, F. Carreño, F. Arrieta-Yañez, and E. Cabrera-Granado, Phys. Rev. A \textbf{78}, 053812 (2008).


\bibitem{Zhu1991} M. Zhu, C. W. Oates, and J. L. Hall, Phys. Rev. Lett. \textbf{67}, 46 (1991).

\bibitem{Yeo2015} M. Yeo, M. T. Hummon, A. L. Collopy, B. Yan, B. Hemmerling, E. Chae, J. M. Doyle, and J. Ye, Phys. Rev. Lett. \textbf{114}, 223003 (2015).

\bibitem{Ravi2025} T. Ravi, R. C. Das, H. V. Bhusane, S. Roy, K. Pandey, arXiv:2505.14301 (2025).

\bibitem{Williams2017} H. J. Williams, S. Truppe, M. Hambach, L. Caldwell, N. J. Fitch, E. A. Hinds, B. E. Sauer, and M. R. Tarbutt, New J. Phys. \textbf{19}, 113035 (2017).

\bibitem{Truppe2017} S. Truppe, H. J. Williams, N. J. Fitch, M. Hambach, T. E. Wall, E. A. Hinds, B. E. Sauer, and M. R. Tarbutt, New J. Phys. \textbf{19}, 022001 (2017).


\bibitem{Breene1961} R. G. Breene, \textit{The shift and shape of spectral lines} (Pergamon, New York, 1961).

\bibitem{Breene1981} R. G. Breene, \textit{Theories of spectral line shapes} (Wiley, New York, 1981).

\bibitem{Breene19812} R. G. Breene, H. R. Zaidi, Physics Today \textbf{34}, 99 (1981).

\bibitem{Allen1975} L. Allen, J. H. Eberly, \textit{Optical resonance and two-level atoms} (Dover, New York, 1975).

\bibitem{Shore1990} B. W. Shore, \emph{The Theory of Coherent Atomic Excitation} (Wiley, New York, 1990).

\bibitem{Citron1977}
M. L. Citron, H. R. Gray, C. W. Gabel, and C. R. Stroud, Jr.,
Phys. Rev. A 16, 1507 (1977).


\bibitem{Bergquist1996} J. C. Bergquist, \textit{Doppler-free spectroscopy (Chapter 13)} (Academic Press, Boulder, 1996). Available online: https://tf.nist.gov/general/pdf/1125.pdf.


\bibitem{Foot2005}
C. J. Foot, \textit{Atomic Physics} (Oxford University Press, 2005) 

\bibitem{Milonni2010}
P. W. Milonni and J. H. Eberly, \textit{Laser Physics} (Wiley, 2010).

\bibitem{Vitanov2001}
N. V. Vitanov, B. W. Shore, L. P. Yatsenko, K. Böhmer, T. Halfmann, T. Rickes, and K. Bergmann, Opt. Commun. \textbf{199}, 117 (2001).

\bibitem{Halfmann2003}
T. Halfmann, T. Rickes, N. V. Vitanov, and K. Bergmann
Opt. Commun. \textbf{220}, 353 (2003).

\bibitem{Harth1995}
A. Harth,
IEEE J. Quant. Electr. \textbf{31}, 894 (1995). 

\bibitem{Roy2011}
C. Roy and S. Hughes,
Phys. Rev. X \textbf{1}, 021009 (2011).

\bibitem{Levine2012}
 J. Levine,
 Spectrochim. Acta B: At. Spectr. \textbf{69}, 61 (2012)

\bibitem{Bettles2020}
R. J. Bettles, M. D. Lee, S. A. Gardiner, and J. Ruostekoski,
Commun. Phys. \textbf{3}, 141 (2020).

\bibitem{Conover2011} C. W. S. Conover, Phys. Rev. A \textbf{84}, 063416 (2011).


\bibitem{Mihov2024} I. S. Mihov and N. V. Vitanov, Phys. Rev. Lett. \textbf{132}, 020802 (2024).

\bibitem{Mihov2023} I. S. Mihov and N. V. Vitanov, Phys. Rev. A \textbf{108}, 042604 (2023).

\bibitem{Florez2013} H. M. Florez, L. S. Cruz, M. H. G. de Miranda, R. A. de Oliveira, J. W. R. Tabosa, M. Martinelli, and D. Felinto, Phys. Rev. A \textbf{88}, 033812 (2013).

\bibitem{Reimer2016} M. E. Reimer, G. Bulgarini, A. Fognini, R. W. Heeres, B. J. Witek, M. A. M. Versteegh, A. Rubino, T. Braun, M. Kamp et al., Phys. Rev. B \textbf{93}, 195316 (2016).



\bibitem{Dreissen2022} L. S. Dreissen, C.-H. Yeh, H. A. F\"urst, K. C. Grensemann, and T. E. Mehlst\"aubler, Nat. Commun. \textbf{13}, 7314 (2022).

\bibitem{Rosen1932} N. Rosen and C. Zener, Phys. Rev. \textbf{40}, 502 (1932).


\bibitem{Robinson1985} E. J. Robinson, J. Phys. B: Atom. Mol. Phys. \textbf{18}, 3687 (1985).

\bibitem{Boradjiev2013} I. I. Boradjiev and N. V. Vitanov, Opt. Commun. \textbf{288}, 91 (2013).


\bibitem{Zhu2022} J. Zhu, X. Laforgue, X. Chen and S. Guérin, J. Phys. B: At. Mol. Opt. Phys. \textbf{55}, 194001 (2022).

\bibitem{Harutyunyan2023} M. Harutyunyan, F. Holweck, D. Sugny, and S. Guérin, Phys. Rev. Lett. \textbf{131}, 200801 (2023).

\bibitem{Peyraut2023} F. Peyraut, F. Holweck, and S. Guérin, Entropy \textbf{25}, 212 (2023).

\bibitem{Kuzmanovic2024} M. Kuzmanović, I. Björkman, J. J. McCord, S. Dogra, and G. S. Paraoanu,  Phys. Rev. Research \textbf{6}, 013188 (2024).

\bibitem{Mccord2025} J. J. McCord, M. Kuzmanović, G. S. Paraoanu, arXiv:2504.16646 (2025).

\bibitem{Boradjiev2013pra}
I. I. Boradjiev and N. V. Vitanov,
Phys. Rev. A \textbf{88}, 013402 (2013).

\bibitem{Saharyan2023} A. Saharyan, B. Rousseaux, Z. Kis, S. Stryzhenko, and S. Guérin, Phys. Rev. Research \textbf{5}, 033056 (2023).

\bibitem{Dridi2024} G. Dridi, X. Laforgue, M. Mejatty, and S. Guérin, Phys. Rev. A \textbf{109}, 062613 (2024).

\bibitem{Mihov20242} I. S. Mihov and N. V. Vitanov, Phys. Rev. A \textbf{110}, 052609 (2024)

\bibitem{Motzoi2009} F. Motzoi, J. M. Gambetta, P. Rebentrost, and F. K. Wilhelm, Phys. Rev. Lett. \textbf{103}, 110501 (2009).

\bibitem{Li2024} B. Li, T. Calarco, and F. Motzoi, npj Quantum Information \textbf{10}, 66 (2024).



\bibitem{Vitanov2009}
N. V. Vitanov, B. W. Shore, and L. P. Yatsenko,
Ukr. J. Phys. \textbf{54}, 53 (2009).


\bibitem{Born1928}
M. Born and V. A. Fock, Z. Phys. A \textbf{51}, 165 (1928).

\bibitem{Berry1990} M. V. Berry, Proc. R. Soc. London A \textbf{429}, 61 (1990).

\bibitem{Joye1993} A. Joye and C.-E. Pfister, J. Math. Phys. \textbf{34}, 454 (1993).

\bibitem{Drese1998} K. Drese and M. Holthaus, Eur. Phys. J. D \textbf{3}, 73 (1998).

\bibitem{IBM} IBM Quantum (2022), URL \texttt{https://quantum-computing.ibm.com}.

\bibitem{Koch2007} J. Koch, T. M. Yu, J. Gambetta, A. A. Houck, D. I. Schuster, J. Majer, A. Blais, M. H. Devoret, S. M. Girvin and R. J. Schoelkopf, Phys. Rev. A \textbf{76}, 042319 (2007).


\end{thebibliography}
\end{document}